\documentstyle[aps,prd,epsfig]{revtex}
\newcommand{\bee}{\begin{equation}}
\newcommand{\ee}{\end{equation}}
\newcommand{\beea}{\begin{eqnarray}}
\newcommand{\eea}{\end{eqnarray}}

\preprint{COLO-HEP-452
}
\begin{document}

\title{Low-lying Fermion Modes, Topology and Light Hadrons in Quenched QCD}
\author{Thomas DeGrand, Anna Hasenfratz}
%
%
\address{
Department of Physics,
University of Colorado, 
        Boulder, CO 80309 USA
}
\date{\today}
\maketitle
\begin{abstract}
We explore the properties of low lying eigenmodes of fermions in the
quenched approximation of lattice QCD. The fermion action is a recently
proposed overlap action
and has  exact chiral symmetry. We find that 
chiral zero-eigenvalue modes are
localized in space and their positions correlate strongly
with the locations 
(as defined through the density of
pure gauge observables) of instantons of the appropriate charge.
Nonchiral modes are also localized with peaks 
which are strongly correlated with
the positions of both charges of instantons.
These correlations slowly die away as the fermion eigenvalue rises.
Correlators made of quark propagators
 restricted to these modes closely reproduce ordinary hadron correlators
  at small quark mass in many channels.
Our results are in qualitative agreement with the expectations of
instanton liquid models.
\end{abstract}
%
%

\section{Introduction}

Is there a particular physical mechanism in QCD which is responsible
for chiral symmetry breaking? If so, what other qualitative
or quantitative features of QCD depend on this mechanism?
The leading candidate for the source of chiral symmetry breaking
is topological (instanton) 
excitation of the gauge field, which couples to the quarks
through the associated fermion zero modes (or near-zero modes, after mixing)
leading to chiral symmetry breaking via the Banks-Casher\cite{ref:BC} relation.
An elaborate phenomenology built on the interactions of fermions
with instantons is said to account for many of the low energy
properties of QCD (for a review, see Ref. \cite{ref:SS,ref:Diak}).

Lattice simulations can in principle address this issue, and indeed this
is a large and active area of research. However, nearly all results,
be they from pure gauge operators or from fermions,
are contaminated by one kind of lattice artifact or another, which
cloud the picture.

The problem is, that typically, pure gauge topological
observables depend on the operator used.
The dominant features of the QCD vacuum seen in any lattice simulation
are just ultraviolet fluctuations, as they would be for any quantum field
theory. To search for instantons (or other objects),
 one must invent operators which
filter out long distance structure from this uninteresting noise.
Some quantities  (like the topological susceptibility in $SU(3)$ gauge
theory) are less sensitive to filtering, but some (like the size distribution
of topological objects) are more so, and most
 results are controversial (see Ref. \cite{ref:negele_rev} for a recent
summary).

Perfect action topological operators \cite{ref:FP,ref:perf_top} offer a mathematically consistent definition for the pure gauge topological charge, but
the implementation of such an operator is prohibitively expensive for QCD.
Even if there is a pure gauge definition of topology,
the situation is still that what is important is what the fermions see.
Here the problem is that until recently, all lattice fermion actions 
were contaminated by chiral symmetry-sensitive artifacts.
The dimension-5 operator which eliminates doubling in Wilson-type
actions breaks chiral symmetry and spreads the
 real eigenmodes of the Dirac operator over a finite range, making the
connection between what would otherwise be zero modes and
instantons problematic.
In principle, even domain wall fermions at finite values of the fifth
dimension would show similar artifacts (though in practice they are much 
reduced).
Staggered fermions generally do not have zero eigenmodes in the presence of
instantons\cite{ref:KSC}: the modes split into imaginary pairs
and one must try to separate the pairs that would be zero modes from the
true non-chiral eigenmodes. At small lattice spacing the chiral
symmetry breaking problems are reduced and some
studies seem to grapple successfully with lattice artifacts,
but the whole situation is rather unsatisfactory\cite{ref:sample}.

The discovery of lattice actions 
which implement an exact chiral symmetry without doubling\cite{ref:reviews}
allows one to revisit these questions in a theoretically clean context.
One explicit realization of such an action is the overlap action
of Neuberger\cite{ref:neuberfer}. It obeys the simplest version of the
Ginsparg-Wilson\cite{ref:GW} (G-W) relation.  
So far there have been many studies of aspects of quenched QCD
with overlap actions built from the usual Wilson fermion action
 in four dimensions\cite{ref:FSU,ref:HJLu,ref:over4d,ref:FSUMORE1,ref:FSUMORE2,ref:FSUMORE3,ref:HJL,ref:LIU}.
 The results we will present here are based on
a new  overlap action recently described by one of us\cite{ref:TOM_OVER}.

The advantage of studying questions relevant to chiral symmetry on the lattice
with an overlap action 
are obvious: the action itself is chiral. Real-eigenvalue
 eigenmodes of the Dirac operator are true chiral zero modes.
No fine tuning of parameters or post-processing of lattice data is
required to do measurements at or close to the chiral limit. All analysis
becomes much simpler.

The main disadvantage of the overlap is its expense:
for the action of Ref. \cite{ref:TOM_OVER}, about a factor of 100 more
costly than the clover action to do any calculation which can be done with 
the clover action. However, overlap calculations can be used to
``validate'' simpler measuring techniques on smaller lattices, and the
simpler techniques can then be used to make measurements on
large lattices.

In this paper we study the low lying eigenmodes of an overlap action.
We observe that these modes show 
a spatially peaked structure: chiral zero modes
correlate with the positions of the appropriately charged
topological objects 
 while nonchiral modes correlate
with the positions of both signs of topological charge.
Next we show that, when the quarks are sufficiently light,
 these eigenmodes contribute in an important way
to ordinary hadron correlators (as used, for example, in a spectroscopy
calculation), just as the instanton phenomenology would expect them
to do so. As the quark mass grows heavier, these modes become less
important even though manifestations of chiral symmetry breaking persist
in spectroscopy and matrix elements.

The study of topology most similar to ours is the investigation of
the deconfined phase of $SU(2)$ gauge theory by Edwards, et. al.
\cite{ref:FSUMORE1}.

In Sec. 2 we give a brief summary of our simulations, and then in Sec. 3
we describe our observations of the properties of low-lying eigenmodes
of the Dirac operator. In Sec. 4 we examine the relevance of these modes
to hadronic correlators. We summarize our results in Sec. 5.

\section{Simulation Parameters}
The overlap action used in these studies\cite{ref:TOM_OVER}
 is built from an action with nearest and
next-nearest neighbor couplings. 
The action uses APE-blocked links\cite{ref:APEblock}:
 Our definition of this blocking is
\beea
V^{(n)}_\mu(x) = &Proj_{SU(3)}\quad \quad  \{
(1-\alpha)V^{(n-1)}_\mu(x) \nonumber  \\
& +   {\alpha \over 6} \sum_{\nu \ne \mu}
(V^{(n-1)}_\nu(x)V^{(n-1)}_\mu(x+\hat \nu)V^{(n-1)}_\nu(x+\hat \mu)^\dagger
\nonumber  \\
& +  V^{(n-1)}_\nu(x- \hat \nu)^\dagger
 V^{(n-1)}_\mu(x- \hat \nu)V^{(n-1)}_\nu(x - \hat \nu +\hat \mu) )\},
\label{APE}
\eea
with  $V^{(n)}_\mu(x)$  projected back onto $SU(3)$ after each step, and
 $V^{(0)}_\mu(n)=U_\mu(n)$ the original link variable. Here we chose $\alpha=0.45$ and perform 10 smearing steps.
Eigenmodes of the overlap Dirac operator $D$ are constructed from eigenmodes
of the Hermitian Dirac operator $H=\gamma_5 D$,
 using an adaptation of
a conjugate gradient algorithm of Bunk et. al.
and  Kalkreuter and Simma\cite{ref:eigen}.
The code makes extensive use of multi-mass Conjugate Gradient matrix
inverters\cite{ref:multimass}.

The data set used in this analysis uses the Wilson gauge action at a coupling
 $\beta=5.9$, or a nominal lattice spacing of
$a \simeq 0.12$ fm. It consists of 20 $12^4$
configurations and 20 $12^3\times 24$ configurations.
 Data collection
took about three months on the Colorado 29 (give or take a few)
  node Beowulf cluster.

On the $12^4$ lattices we
 calculated the ten lowest-eigenvalue eigenmodes on each configuration
(more properly: the ten smallest eigenvalue modes of $H^2$ in the chiral
sector of the  minimum eigenvalue, and reconstructed the degenerate eigenstate
of opposite chirality when one was present).
On one configuration we found the lowest twenty
 eigenmodes. On the $12^3\times 24$
lattices we found the lowest twenty eigenmodes.

\begin{figure}[thb]
\begin{center}
\epsfxsize=0.7 \hsize
\epsffile{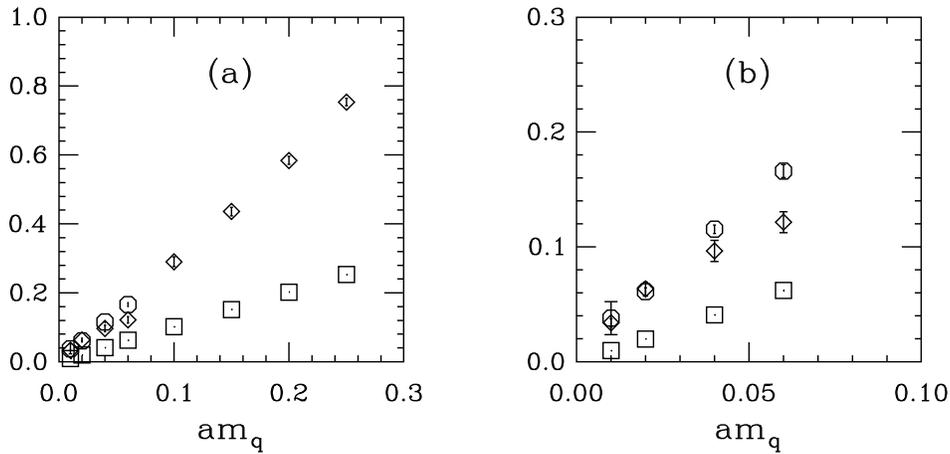}
\end{center}
\caption{
Squared pion mass (diamonds and octagons)
 and quark mass from the PCAC relation (squares)
 from the planar overlap action. Panel (b) merely blows up the
small quark mass data of panel (a).
The four lighter mass values are from the $12^3\times 24$ lattices.
Diamonds show pion masses extracted from a fit to the pseudoscalar-pseudoscalar
correlator, and octagons show the pion mass extracted from
a fit to the difference of the pseudoscalar-pseudoscalar and scalar-scalar
correlators. See Sec. 4 for further discussion.}
\label{fig:mpi2mq}
\end{figure}

We have done an exploratory spectroscopy calculation on these data sets,
at four values of the light quark mass.
We also computed the spectrum at four heavier quark masses on the $12^4$ set.
We show two sets of results, simply
to demonstrate how un-exceptional our measurements are.
Fig. \ref{fig:mpi2mq} shows the squared pion mass (from two
correlators) and the quark mass
(from the PCAC relation). As expected with a chiral action, no
additive mass renormalization for the quark mass is observed.
 (The squared pion mass extracted from the correlator of two
pseudoscalar currents does not extrapolate to zero linearly at small
 bare quark mass, though the mass extracted from the
difference of the scalar correlator and the pseudoscalar correlator does.
 This might be a quenched
 approximation-finite volume artifact, as we discuss in Sec. 4.)

\begin{figure}[thb]
\begin{center}
\epsfxsize=0.5 \hsize
\epsffile{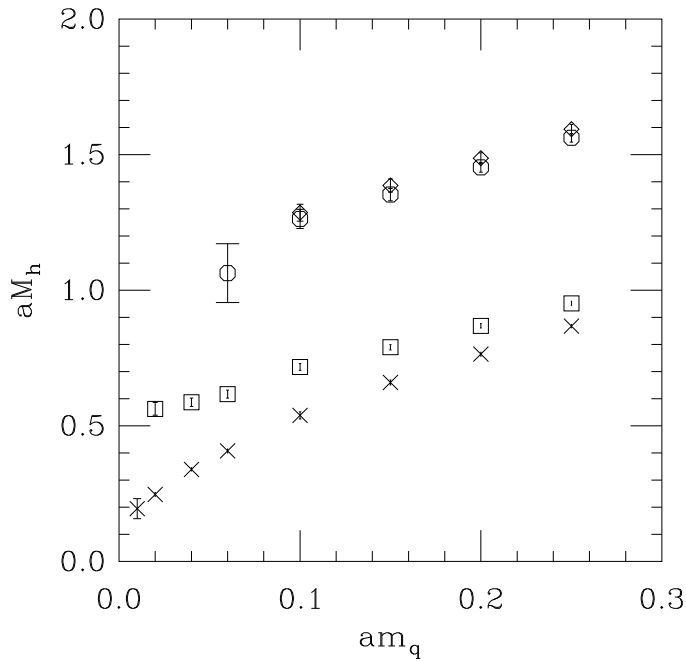}
\end{center}
\caption{
Spectroscopy from  the planar overlap action:
crosses, pseudoscalar mesons; squares, vector mesons; octagons, nucleons;
diamonds, deltas.
The four lighter mass values are from the $12^3\times 24$ lattices.
}
\label{fig:spec}
\end{figure}

Results for spectroscopy are shown in Fig. \ref{fig:spec}. With the
small data set, the noise in the baryon channels becomes too large
for a stable mass fit at small quark mass.
 Making a naive
 extrapolation
of the rho mass to the chiral limit and fixing the lattice spacing from
the physical value of its mass  would indicate a lattice spacing of
about 0.13 fm, as opposed to a value of 0.11 fm from the Sommer parameter,
 using
the interpolating formula of Ref. \cite{ref:precis}.

\begin{figure}[thb]
\begin{center}
\epsfxsize=0.5 \hsize
\epsffile{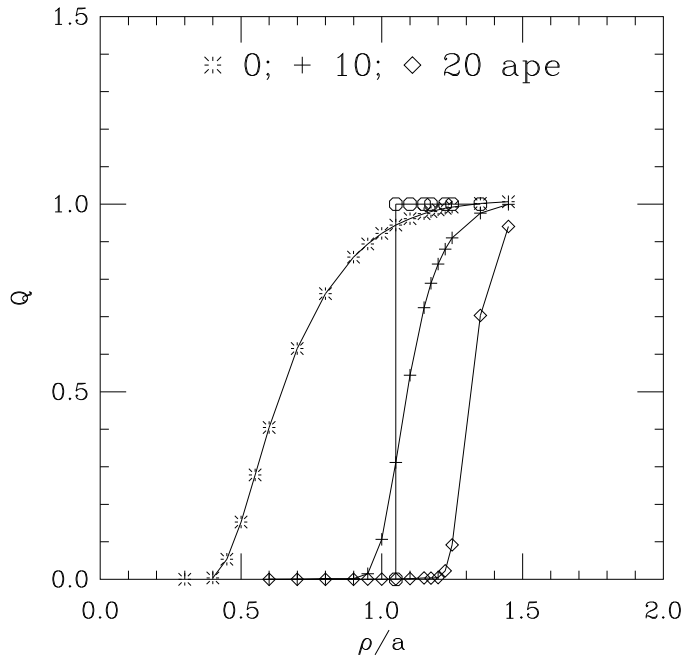}
\end{center}
\caption{Comparison of topological charge on a set of smooth single instanton
configurations of varying instanton radius $\rho$. Octagons show the number
of zero modes of the overlap fermion, while the other symbols show the 
topological charge measured by the pure gauge observable with various amounts
of APE-smearing.
}
\label{fig:QsmoothapeT}
\end{figure}

We measured the topological charge density (which we denote as $Q(x)$)
using an operator described in our previous work\cite{ref:COLO_INST}.
This operator is built of a sum of two perimeter-ten loops,
twisted in four dimensions to provide a lattice approximation to
${\rm{Tr}}F_{\mu\nu}(x) \tilde F_{\mu\nu}(x)$. The links in the loops
are built of APE-smeared links, and the operator can be tuned by
the choice of $\alpha$ (0.45 in this study) and $N$ (mostly set to 10).
As a demonstration of the efficiency of the operator, we examined a set
of smooth single instanton configurations, varying the radius $\rho$
 of the instanton.
Our results are shown in Fig. \ref{fig:QsmoothapeT}. Without smearing the 
topological charge  $Q=\int d^4xQ(x)$ decreases slowly from one to zero 
over a wide range
of instanton radii as the instanton disappears from the lattice, while
at large smearing
levels
it cuts off sharply.
The fermion spectrum should (and does) possess a single, chiral zero
mode for as long as the fermion couples to the instanton. When the
 instanton radius shrinks to some minimum value, the fermion cannot see the
instanton and the zero mode disappears. This behavior is also plotted
in Fig. \ref{fig:QsmoothapeT}. We see that the $\alpha=0.45$,
$N=10$ gauge observable has a
similar response to that of our fermion
 action (which is also built of (0.45,10)
blocked links).

\section{Properties of Low-lying Eigenmodes}

\begin{figure}[thb]
\begin{center}
\epsfxsize=0.4 \hsize
\epsffile{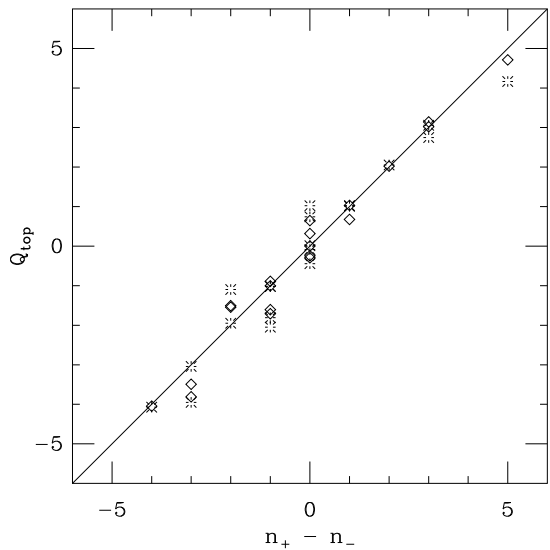}
\end{center}
\caption{Topological charge as measured by the pure gauge observable versus
the number of fermion zero modes for our ensemble of configurations. Diamonds
show 10 APE-smearings for the gauge observable, bursts, 20 smearing steps.
}
\label{fig:Qvsn}
\end{figure}

We begin by comparing the bulk quantities of our configurations.
The Atiyah-Singer index theorem  relates the topological 
charge and the number of exact fermionic
zero modes of a configuration,
$Q_{top}=n_+-n_-$. This relation does not hold exactly in our
case as our definition of $Q_{top}$ is not perfect. Yet
the number of zero modes  and the net topological
charge are strongly correlated, as seen in Fig. \ref{fig:Qvsn}.

An accurate measurement of the topological susceptibility
 $\chi = \langle Q^2 \rangle/V$,
requires  large statistics.
 Nevertheless, the good agreement of the topological charge with the number
of fermion zero modes justifies gauge measurements of 
$\chi$. Nearly all groups\cite{ref:negele_rev} measure a value close to
$\chi^{1/4}=180-200$ MeV.
With our limited data set (on the $12^4$ lattices)
 we find $\chi^{1/4}=205 \pm 15$ MeV.

Next, we examine spatial distributions of the eigenmodes themselves.
To do this, we compute the local chiral density
 $\omega(x) = \langle \psi(x) | \gamma_5 | \psi(x) \rangle$ for each
eigenmode $\psi(x)$ of the Dirac operator $D$. A glance at contour
plots of this quantity reveals a rich structure of bumps. A pattern recognition
program which recognizes the bumps reveals that the locations of maxima
 of different eigenmodes
are strongly correlated, and that they are strongly correlated with the
 positions of maxima of $Q(x)$, the pure gauge observable. The (chiral)
zero modes are spread over maxima of $Q(x)$ which have the opposite sign
($\omega(x)$ is negative on instantons where $Q(x)$ is positive.)
The volume integral of $\omega(x)$ is zero for the nonchiral modes.
These modes have both positive and negative peaks.
The locations of these peaks correlate with the appropriate sign peaks
of $Q(x)$.  This suggests fitting the chiral density in terms of
 single instanton modes
\bee
\omega(x)=\sum_i c_i \omega^0_\rho(x-x_i) +\eta(x),
\ee
where $\omega^0_\rho(x)$ is the chiral density of a single instanton of radius 
$\rho$ and location $x=0$,
  and  $\eta(x)$ represents the part of the chiral density not 
associated with instantons. The coefficients $c_i>0$ for anti-instantons, 
$c_i<0$ for instantons.

\begin{figure}[thb]
\begin{center}
\epsfxsize=0.4 \hsize
\epsffile{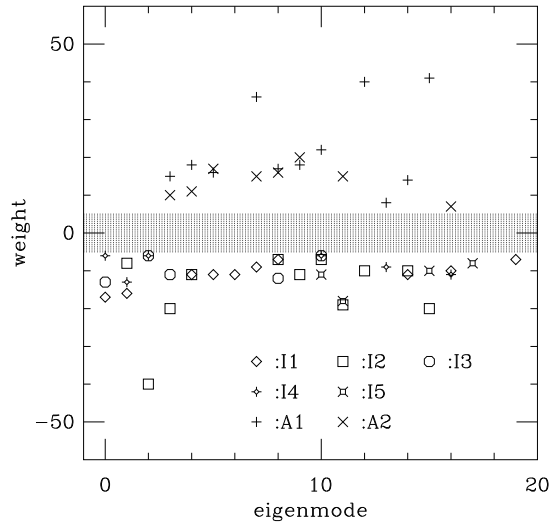}
\end{center}
\caption{
The weight 
with which the most prominent
5 instantons and 2 anti-instantons couple to the first 20 eigenmodes of
a typical $12^4$ configuration.
In the shaded region the background fluctuations are too strong
to distinguish topological modes; therefore, that region
in the graph is excluded.
Observe  that the first three modes (the chiral modes)
couple only to the instantons, while the rest of the modes are distributed
over both instantons and anti-instantons.
}
\label{fig:eig_vs_weight}
\end{figure}

We consider one $12^4$ configuration in detail.
This configuration has topological charge $Q=3$  according to both
the gauge observables and the fermionic definition.
After 10 APE steps the instanton pattern recognition code identifies 
7 instantons and 5 anti-instantons on the gauge configuration. 
After an additional 10 APE steps
the gauge configuration is considerably smoother, and we can identify only 
4 of the instantons and 2 of the anti-instantons, though 
the integrated topological charge is still 3.
Obviously the pattern recognition algorithm misinterprets some object(s). 

The chiral density function $\omega(x)$ also peaks at the same locations. It 
couples strongly to all the objects found after 20 APE steps and picks up
one of the anti-instantons found only after 10 smoothing steps. With smaller
weights it peaks at other locations as well, possibly identifying nearby
pairs. The different eigenmodes couple to the different topological objects
with varying weights. In Fig. \ref{fig:eig_vs_weight} we show the weight 
relative to the background fluctuations, 
\bee
w_i={c_i \over \sqrt{|\eta|^2}}
\ee
 with which the most prominent
5 instantons and 2 anti-instantons couple to the first 20 eigenmodes.
In order to identify a mode we require that its weight is at least five
times above background, $w_i > 5$,
 therefore the shaded region in the graph is excluded.
Observe from the figure that the first three modes (the chiral modes)
couple only to the instantons, while the rest of the modes are distributed
over both instantons and anti-instantons.
Most modes couple with weights 10-20 to several
topological objects, but occasionally the mode overwhelmingly 
couples to a single topological object. 

There is no qualitative difference between the lowest and highest non-chiral
modes: they couple to the same set of topological objects with slightly 
decreasing magnitude. The volume of this configuration is about 4 fm$^4$, so
according to the instanton-liquid model one would expect around 4 well defined
objects. Indeed the pattern recognition code identifies about half a dozen
topological objects, yet every one of
  the lowest 20 eigenmodes show strong coupling
to them and there is no reason to believe that the situation will change if
we consider the next 10 or 20 eigenmodes. They all couple to the same set of
 topological modes in addition to non-topological, "spin-wave" modes. As the 
eigenvalue increases, the coupling to the topological modes decreases while
the spin wave mode coupling increases, until it is no  longer 
possible to actually
separate the topological modes from the background. 
 The $12^3 \times 24$ configurations
show a similar pattern but with more topological objects, as it is expected
in a larger volume.

To further quantify these observations, we construct the autocorrelation
 function of chirality
\bee
C_{\omega,\omega}(r)= {1\over V}\int d^3x \omega(x)\omega(r+x)
\ee
and the correlation function of chirality with topological charge
\bee
C_{\omega,Q}(r)= {1\over V}\int d^3x \omega(x)Q(r+x).
\ee

\begin{figure}[thb]
\begin{center}
\epsfxsize=0.4 \hsize
\epsffile{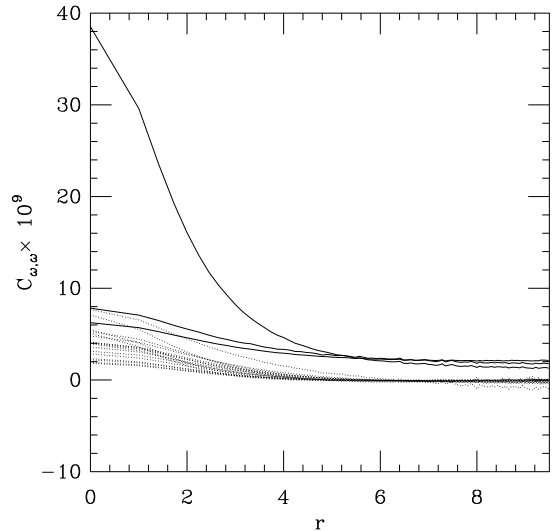}
\end{center}
\caption{ Chirality correlation function for a $12^4$
 configuration with 3 zero modes
(shown as solid lines),
showing the additional 17 lowest non-chiral mode pairs
 (shown as dotted lines). The height of
the correlator at the origin decreases more or less
monotonically with the size of the eigenvalue of the mode.
}
\label{fig:cheigc5all}
\end{figure}

\begin{figure}[thb]
\begin{center}
\epsfxsize=0.4 \hsize
\epsffile{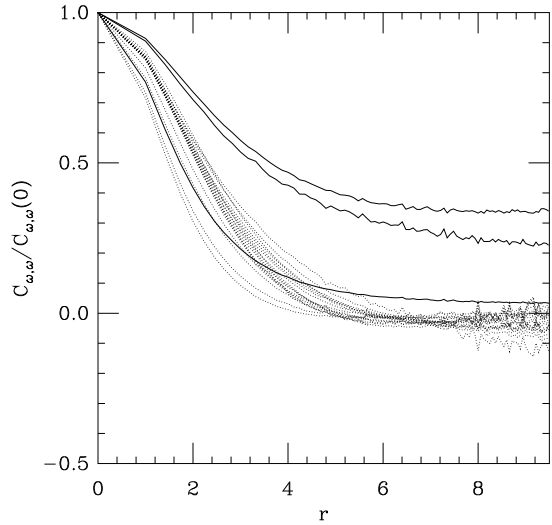}
\end{center}
\caption{The correlator of Fig, \protect{\ref{fig:cheigc5all}}, scaled
by the value of the correlator at the origin. Chiral modes are shown as
solid lines, while the nonchiral modes are shown as dotted lines.
}
\label{fig:cheigc5rescaled}
\end{figure}

Let us first consider $C_{\omega,\omega}(r)$ on the same $Q=3$ configuration
as above, shown in Fig. \ref{fig:cheigc5all}. 
We show the autocorrelator for each of the lowest 20
 modes, chiral and nonchiral. The peak at small $r$
 indicates that the chirality
is localized. The amount of localization of a mode dies away slowly
as the eigenvalue of the mode increases.
(The autocorrelation function of the zero modes does not fall to zero because
$\omega(x)$ integrates to unity, while $\int \omega(x)=0$ for the nonchiral
modes.)
However, the size of the localized region does not depend too much on the
magnitude of the eigenvalue. We can see that by normalizing the autocorrelator
by its value at the origin, in Fig. \ref{fig:cheigc5rescaled}.

\begin{figure}[thb]
\begin{center}
\epsfxsize=0.4 \hsize
\epsffile{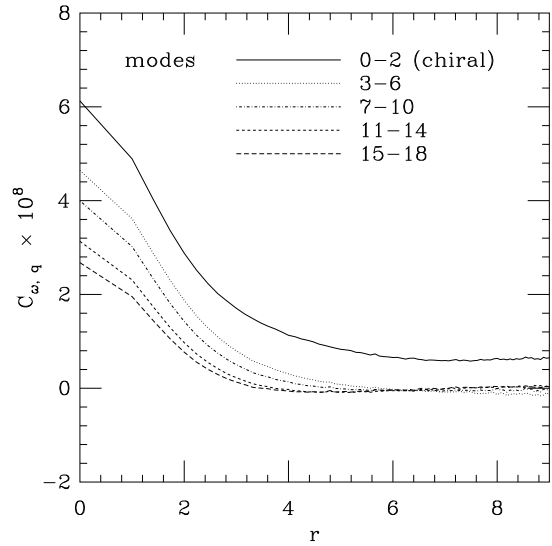}
\end{center}
\caption{Mixed topological charge-chirality correlation function for a typical
configuration. The correlators for the three chiral modes are averaged,
as are the correlators for groups of nonchiral modes (organized by increasing
fermion eigenvalue).
}
\label{fig:chtopeigc5allaver}
\end{figure}

\begin{figure}[thb]
\begin{center}
\epsfxsize=0.4 \hsize
\epsffile{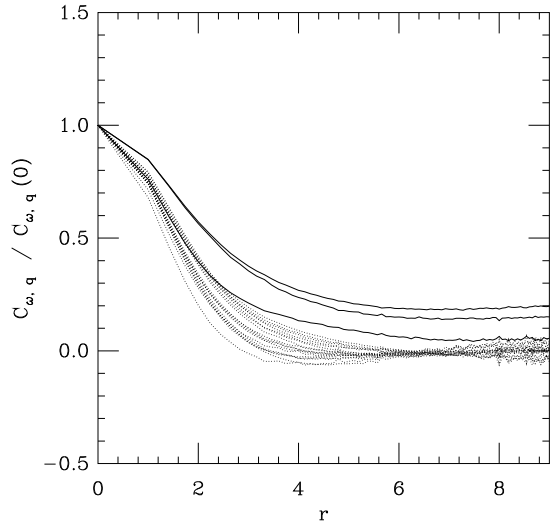}
\end{center}
\caption{Mixed topological charge-chirality correlation function for  the
configuration of Fig. \protect{\ref{fig:chtopeigc5allaver}}, scaled
to its value at the origin. Chiral modes are solid, nonchiral ones, dotted.
}
\label{fig:chtopeigc5allrescaled}
\end{figure}

Similar features are seen for the mixed correlator. Where
 the topological charge is large, there also $\omega(x)$ peaks.
Higher modes gradually decorrelate with topological charge.
Nevertheless, the size of the correlated region depends only weakly on
the eigenvalue.
Compare Figs. \ref{fig:chtopeigc5allaver} and
\ref{fig:chtopeigc5allrescaled}.

\begin{figure}[thb]
\begin{center}
\epsfxsize=0.4 \hsize
\epsffile{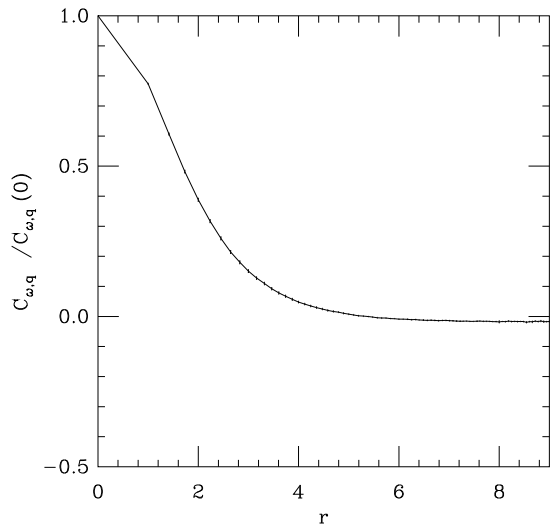}
\end{center}
\caption{Mixed topological charge-chirality correlation function
averaged over all the nonchiral modes on all $12^4$
 configurations in the data set.
}
\label{fig:chtopeigallnonchiralrescaledaver}
\end{figure}

\begin{figure}[thb]
\begin{center}
\epsfxsize=0.4 \hsize
\epsffile{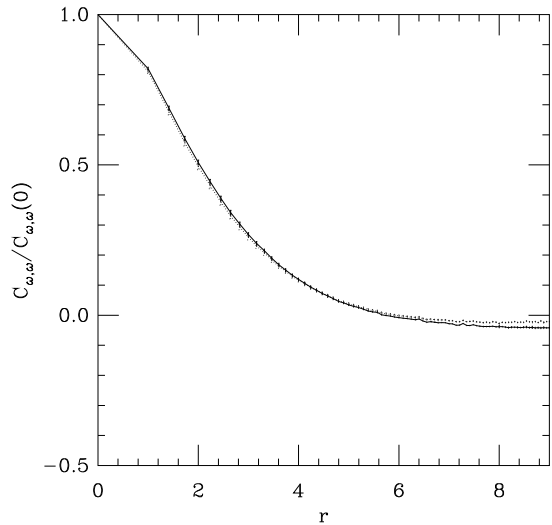}
\end{center}
\caption{Chirality autocorrelation function
averaged over all the nonchiral modes on all configurations in the data set.
 The lower curve corresponds to $12^4$ lattices,
the upper one to $12^3 \times 24$ lattices.
}
\label{fig:cheigcallnonchiralrescaledaver}
\end{figure}

Now we combine all the lattices in our data set and average our correlation
functions over all the nonchiral modes. The  correlation function for the
chirality with the topological charge
 and the autocorrelation function of the chirality  are shown in Figs.
\ref{fig:chtopeigallnonchiralrescaledaver}
and \ref{fig:cheigcallnonchiralrescaledaver}.
(There really are error bars in these pictures!)
 The chirality autocorrelation function is slightly broader than the
 mixed correlator, which is not surprising
as the fermionic zero mode wave function falls off slower than the topological
charge density.
 Fig. \ref{fig:cheigcallnonchiralrescaledaver} actually has
two curves: one corresponds to the $12^{4}$, the other to the $12^3\times 24$
data set. The two curves are almost indistinguishable except at large distances
where, due to the periodicity of the lattice, the $12^4$ curve falls
a bit below  the other.

All of these features are exactly what one would expect based on
the instanton liquid model of the QCD vacuum\cite{ref:SS}:
The (degenerate) chiral modes sit on the appropriate charge 
instantons, coupling to all of them. The nonchiral modes are made of
a superposition of peaks, each peak centered on an instanton or an 
anti-instanton, and interpolating among them.

\begin{figure}[thb]
\begin{center}
\epsfxsize=0.4 \hsize
\epsffile{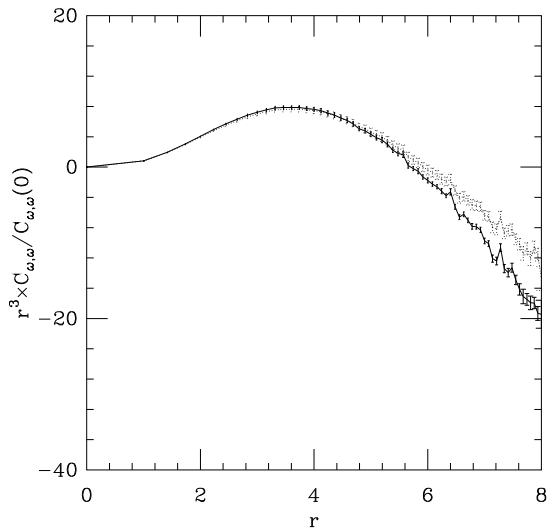}
\end{center}
\caption{The same data as
 Fig. \protect{\ref{fig:cheigcallnonchiralrescaledaver}}, but scaled by
$r^3$ to expose the large distance tail.
The solid curve corresponds to $12^4$ lattices,
 the dotted one to $12^3 \times 24$ lattices.
}
\label{fig:cheigcallnonchiralr3aver}
\end{figure}

If the chirality correlator is ``lumped,'' and interpolates between
opposite sign peaks sitting on instantons and anti-instantons, one would
also expect that $C_{\omega,\omega}(r)$ would go negative at large $r$,
as an instanton peak (one sign of chirality) would anti-correlate
 with a nearby 
anti-instanton peak (opposite chirality).
 Fig. \ref{fig:cheigcallnonchiralr3aver} multiplies the autocorrelator by
the phase space factor of $r^3$ and exposes this behavior.
Similar negative correlation was observed using the pure gauge topological
 charge operator in Ref. \cite{ref:molecules}.

\begin{figure}[thb]
\begin{center}
\epsfxsize=0.5 \hsize
\epsffile{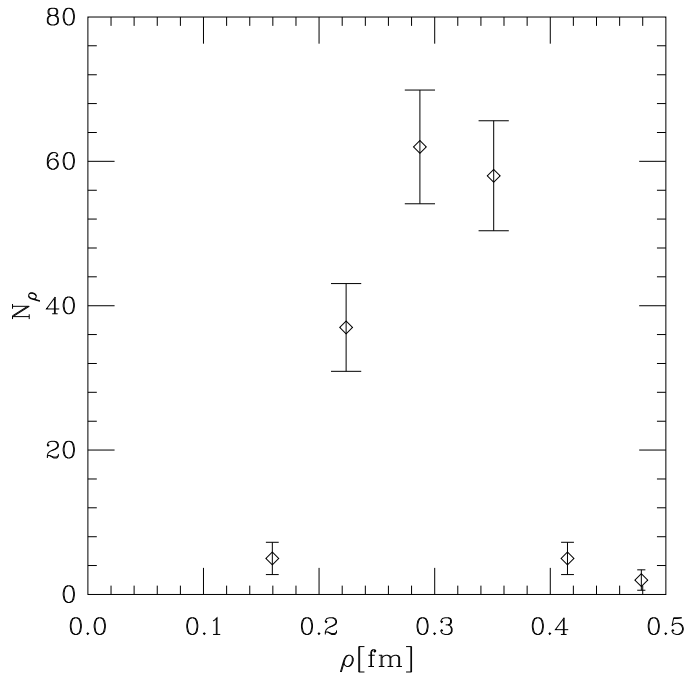}
\end{center}
\caption{Instanton number density vs size, extracted from the fermion
chirality density function, and converted to physical units using
a nominal lattice spacing of $a=0.12$ fm.
}
\label{fig:drho}
\end{figure}

Finally, we can determine a size distribution of the topological objects
seen by the fermions. We do this by identifying peaks in the distribution of
$\omega(x)$, and fitting the peak shape to the shape expected from
a fermionic zero mode. The radius of the peak is directly related to
the instanton radius $\rho$. This is basically the same procedure that is used
to identify instantons with the pure gauge local charge distribution $Q(x)$.
It suffers from the same limitations--a bump must stand out from the background
high enough to be seen, and must not die away too quickly. The first
constraint means that large objects will be lost (large $\rho$ instantons
have a flat profile); the second constraint means that
 small instantons will also not be seen.
The resulting distribution is shown in Fig. \ref{fig:drho}, where we have
converted our result to physical units using a 
nominal lattice spacing of $a=0.12$ fm. The distribution peaks around
$\rho=0.3$ fm.
This value is quite similar to that from a pure gauge calculation by
one of us\cite{ref:COLO_INST}, considerably smaller than that of
two other pure gauge calculations\cite{ref:DEF,ref:SMT}, and quite consistent
with the expectations of instanton liquid
 phenomenology\cite{ref:SS,Shuryak:1995pv}.

\section{What Do These Eigenmodes Do?}
In the last section we demonstrated that the low lying eigenmodes of the
Dirac operator are strongly correlated with the topological structure in the
QCD vacuum. But are these modes relevant to any physical processes?
Phenomenological instanton models predict that light quarks propagate in
the vacuum by hopping between oppositely charged topological objects,
and therefore the lightest hadron propagators are dominated by
 the low lying fermion modes.
There are three obvious classes of observables to investigate:
ordinary hadronic correlators, point-to-point correlators,
and disconnected (hairpin) diagrams.
We have chosen to focus on the first class, since they
directly probe the behavior of QCD at large distances and are relevant to
most measurements of hadronic matrix elements.
(The latter observables are obviously targets for future study.)

We constructed sets of quark propagators on the $12^3 \times 24$
data set at several values of the
quark mass via a usual calculation, and built complementary sets
of propagators in which the quark propagator was approximated by a mode
sum over the lowest N (2 to 20) modes. In both cases we used
a Coulomb-gauge Gaussian source and projected the point sink
 of the correlator onto
zero momentum. We will compare the correlators at lattice quark mass
$am_q=0.01$, 0.02, 0.04, and 0.06, corresponding to a pseudoscalar/vector
meson mass ratio of about 0.34, 0.50, 0.61 and 0.64. 

Approximating the quark propagator by a truncated mode sum is obviously
 uncontrolled, and we would not advocate doing it in a real calculation.
However, it serves as a naive realization of an instanton liquid model.

First, we must make a digression to discuss the effect of the exact zero modes.
There are hadron correlators which
are sensitive to zero modes and ones which are not\cite{ref:BNL}.
 Since the zero modes are chiral,
the zero mode quark propagator is proportional to $(1\pm\gamma_5)$ and
inversely proportional to $1/m_q$. Because of its Dirac structure it
couples to the pseudoscalar-pseudoscalar (Ps-Ps) and scalar-scalar (S-S)
 correlators.
 At small quark mass it makes an ever larger contribution to these
 correlators, dominating them in the chiral limit in finite volumes.
  This contribution
is a quenched approximation finite-volume lattice artifact: quenched,
because the zero mass limit of full QCD has no zero modes, and a finite-volume
effect because the number of zero modes compared to the number of nonchiral
modes is volume-dependent. 
For example, in the Gell-Mann, Oakes, Renner (GMOR)
 relation, the volume-averaged point-to-point Ps-Ps
correlator $\sum_x\langle \pi(x) \pi(0)\rangle/V$ has a contribution
$Q/(m_q^2 V)$ in addition to a volume-independent piece from
 the nonchiral modes. This piece actually scales like
$1/\sqrt{V}$ because $\langle Q^2\rangle\simeq V$.
The axial current autocorrelator can have ``mixed'' contributions
 (one quark propagating through a zero mode and the other quark through
nonzero modes), and the zero modes can be completely decoupled from
the pseudoscalar channel by considering the difference of a
Ps-Ps and S-S correlator.

Since the overall $1/m_q$ coefficient allows the
zero modes to dominate some correlators, a conventional extraction of
 masses from exponential decay
when $m_q$ is much smaller than the eigenvalue of the first nonzero mode
 would give a prospective pion mass which
is independent of the quark mass (because it is related to the
correlations inherent in the zero modes).
 We believe that this artifact might be related to
the flattening of the pion mass seen at  in the Ps-Ps
correlator at $am_q=0.04$ and 0.06
 in Fig. \ref{fig:mpi2mq}. From this correlator, the lowest four points
give a lattice squared pion mass of $(am_\pi)^2=0.018(4)$
 at zero bare quark mass.
The pion mass extrapolated from the
 difference between Ps-Ps and S-S 
correlators, shown in the figure  as octagons, extrapolates to
$(am_\pi)^2=0.008(5)$ at zero quark mass.

The vector current,  $\gamma_i\gamma_5$ axial
current, and the nucleon and Delta all have no coupling to the zero modes.

We find from our data that  at small quark mass  the low modes saturate the
hadron correlators at large time separations. As more modes are added, the
saturation extends to lower and lower times.  This effect decreases
as the quark mass rises.

\subsection{Pseudoscalar and Scalar Correlators}

\begin{figure}[thb]
\begin{center}
\epsfxsize=0.8 \hsize
\epsffile{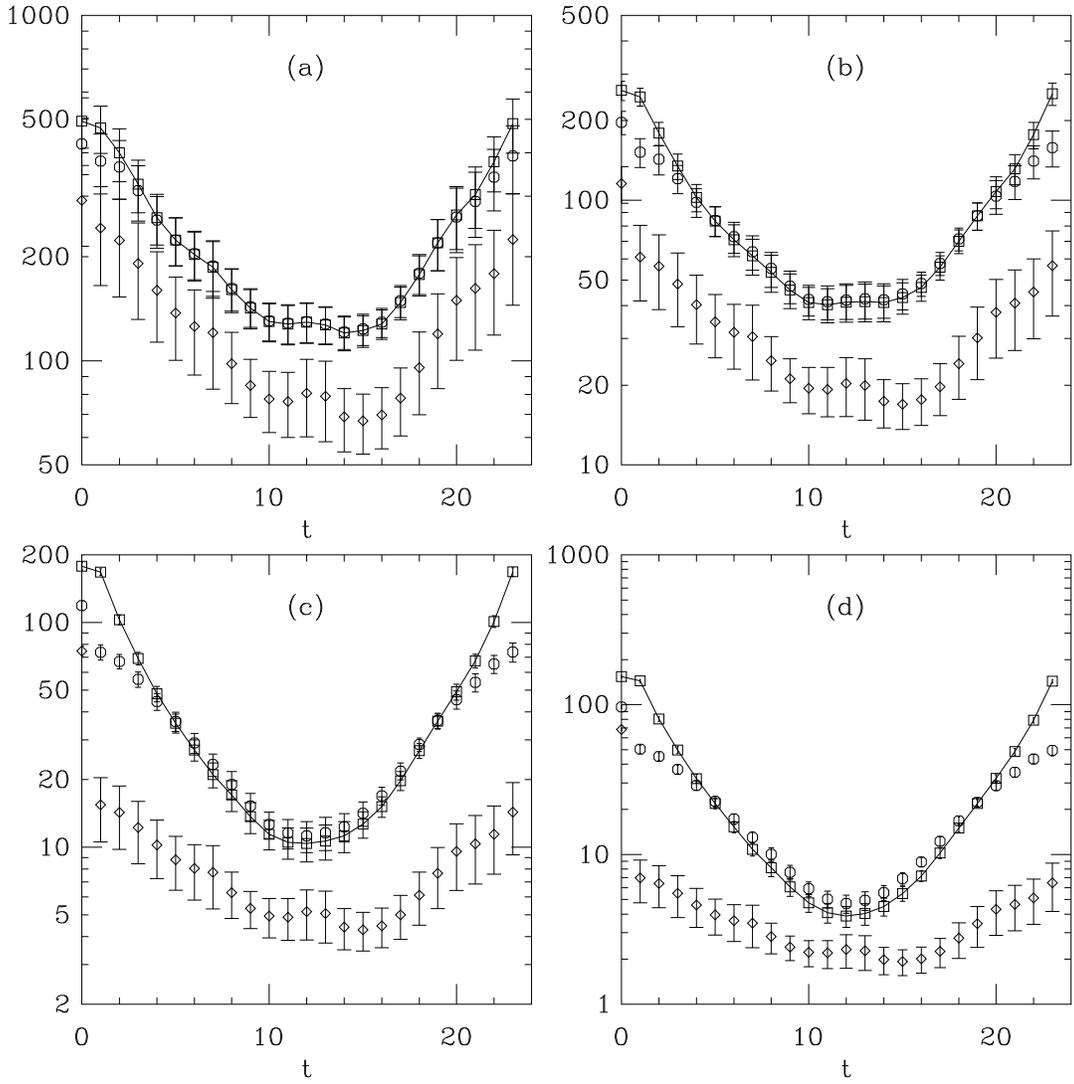}
\end{center}
\caption{
Saturation of the pseudoscalar correlator by  low-lying eigenmodes of $H(0)^2$.
 (a) $am_q=0.01$ ($\pi/\rho\simeq 0.34$);
 (b) $am_q=0.02$ ($\pi/\rho\simeq 0.50$);
 (c) $am_q=0.04$ ($\pi/\rho\simeq 0.61$);
 (d) $am_q=0.06$ ($\pi/\rho\simeq 0.64$).
 Squares (connected by lines) show the full hadron correlator.
Octagons show the contribution from the lowest 20 modes.
 Diamonds show the contribution from the zero modes.
}
\label{fig:comppseudo}
\end{figure}

\begin{figure}[thb]
\begin{center}
\epsfxsize=0.8 \hsize
\epsffile{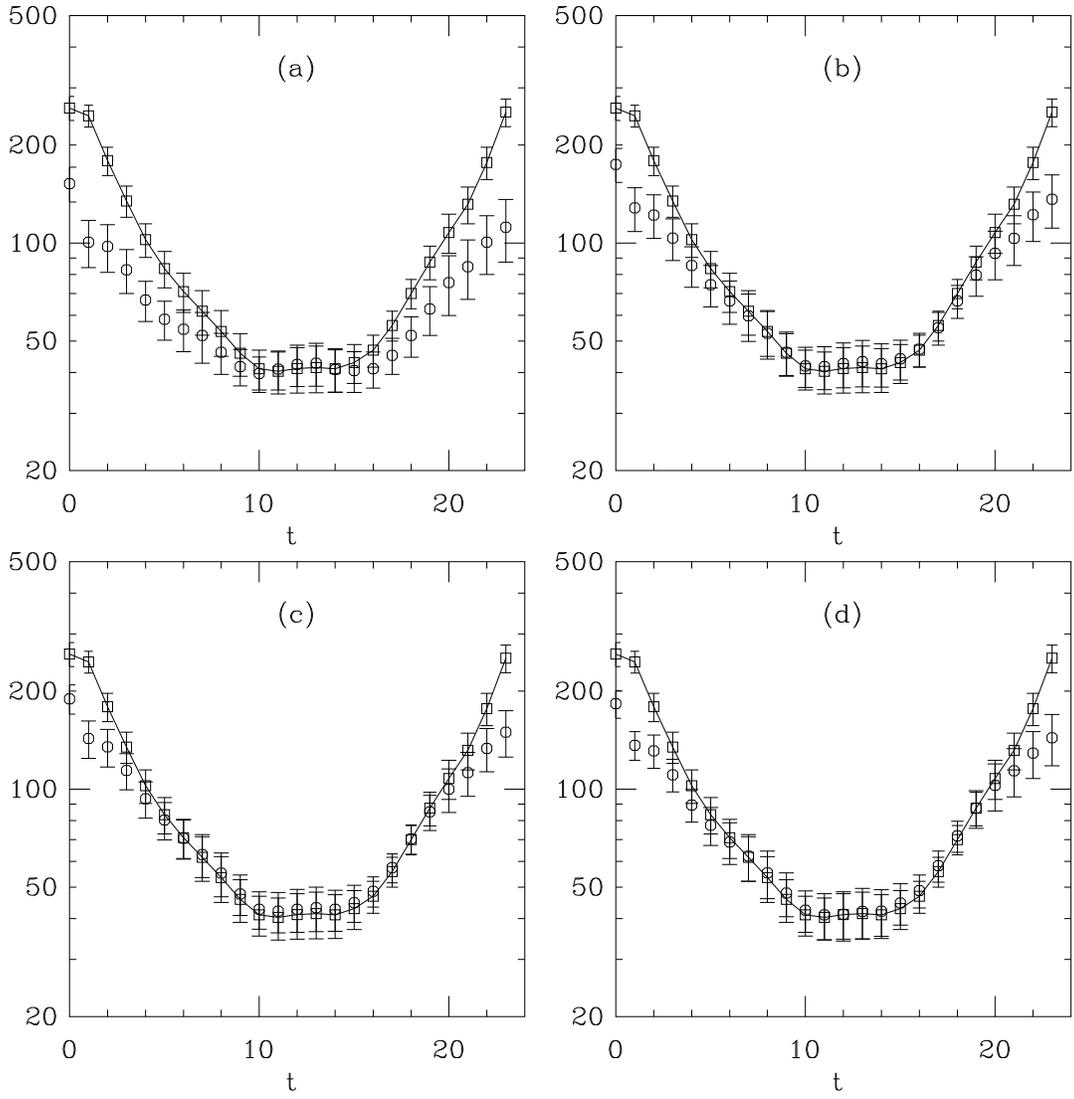}
\end{center}
\caption{
Saturation of the pseudoscalar correlator by  low-lying eigenmodes of $H(0)^2$.
at $am_q=0.02$ ($\pi/\rho\simeq 0.50$).
 Squares (connected by lines) show the full hadron correlator.
Octagons show the contributions from the lowest $N$ modes, where
in (a) $N=4$,
in (b) $N=8$,
in (c) $N=12$,
in (d) $N=16$.
}
\label{fig:compmpseudo}
\end{figure}

\begin{figure}[thb]
\begin{center}
\epsfxsize=0.8 \hsize
\epsffile{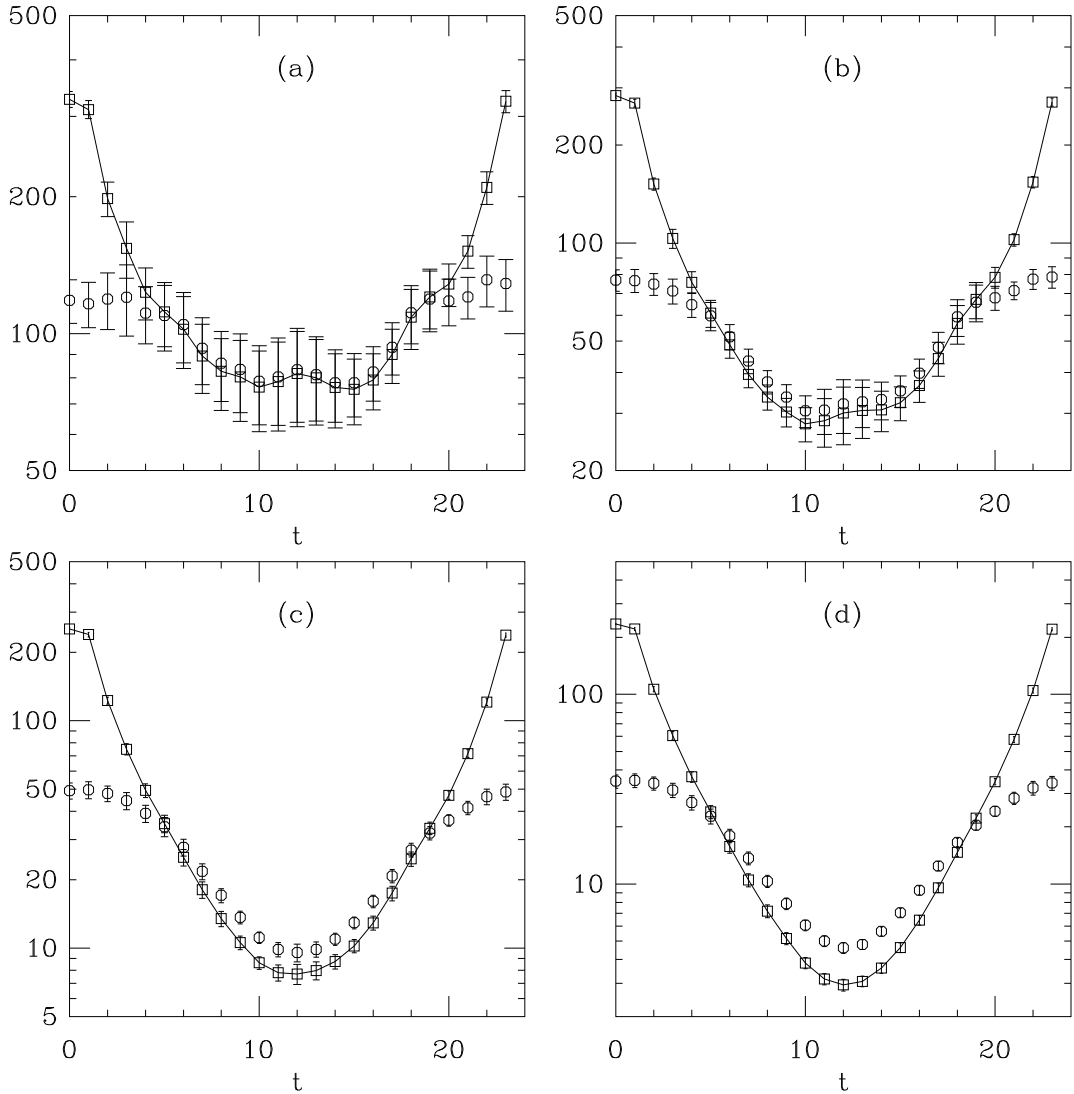}
\end{center}
\caption{
Saturation of the difference of pseudoscalar and
scalar correlators by  the
 20 low-lying eigenmodes of $H(0)^2$.
 Squares (connected by lines) show the full hadron correlator.
Octagons show the contribution from the lowest 20 modes.
 (a) $am_q=0.01$ ($\pi/\rho\simeq 0.34$);
 (b) $am_q=0.02$ ($\pi/\rho\simeq 0.50$);
 (c) $am_q=0.04$ ($\pi/\rho\simeq 0.61$);
 (d) $am_q=0.06$ ($\pi/\rho\simeq 0.64$).}
\label{fig:comppss}
\end{figure}

The most dramatic effects are seen in the pseudoscalar and scalar channels.
  At our  lowest mass
(pseudoscalar/vector meson mass ratio of about 0.34)
 the saturation seems complete,
and at higher masses it is less so.
In Fig. \ref{fig:comppseudo} we compare the full propagator and the $N=20$
truncated propagator for the pseudoscalar channel. Twenty eigenmodes
saturate the propagator even at our heaviest quark mass, $am_q=0.06$.
While the chiral modes themselves do not saturate the propagator
(also shown in Fig. \ref{fig:comppseudo}), at
low mass, only a few modes are needed.
Results for varying the number of modes used to saturate the
pseudoscalar correlator at $am_q=0.02$ are shown in Fig.
   \ref{fig:compmpseudo}.
 At $am_{q}=0.02$ 8-12 modes saturate the
propagator. This is in agreement with instanton models that suggests that the
number of relevant modes is about the same as the number of instantons on the
lattice. With an instanton density of one per fm${}^4$ on our
$12^3 \times 24 \approx 8$ 
fm${}^4$ lattices we indeed expect about 8 instantons per lattice. Results
for the other quark masses are similar though the number of modes needed to
saturate the propagator rises to about 12-16 modes at $\pi /\rho \simeq 0.64$.
Saturation of the pseudoscalar-scalar difference
correlator (to which zero modes do not contribute) is shown in Fig.   
 \ref{fig:comppss}.

We also looked at the simplest quantities related to chiral symmetry
which we can extract from our data, the PCAC quark mass and the pseudoscalar
decay constant.
In an ordinary calculation the ratio of the axial vector matrix element
$A_0 = \bar \psi \gamma_0\gamma_5 \psi$ to the
pseudoscalar matrix element gives the so-called PCAC quark mass
\bee
 2m_q = {{ \sum_{\vec y}\langle \partial_0 A_0(\vec y, y_0) C(0,0) \rangle}
\over { \sum_{\vec y}\langle P(\vec y, y_0) C(0,0) \rangle}}
\label{RHO}
\ee
and the same numerator is used in the lattice measurement of the
 pseudoscalar decay constant $f_{PS} \simeq \langle \pi |A_0 |0\rangle$.
Overlap fermions satisfy the GMOR relation mode by mode.
This is not the case for the PCAC relation, although one would
expect that a chiral theory would also respect it. What do low eigenmode
truncations give for these quantities?

\begin{figure}[thb]
\begin{center}
\epsfxsize=0.8 \hsize
\epsffile{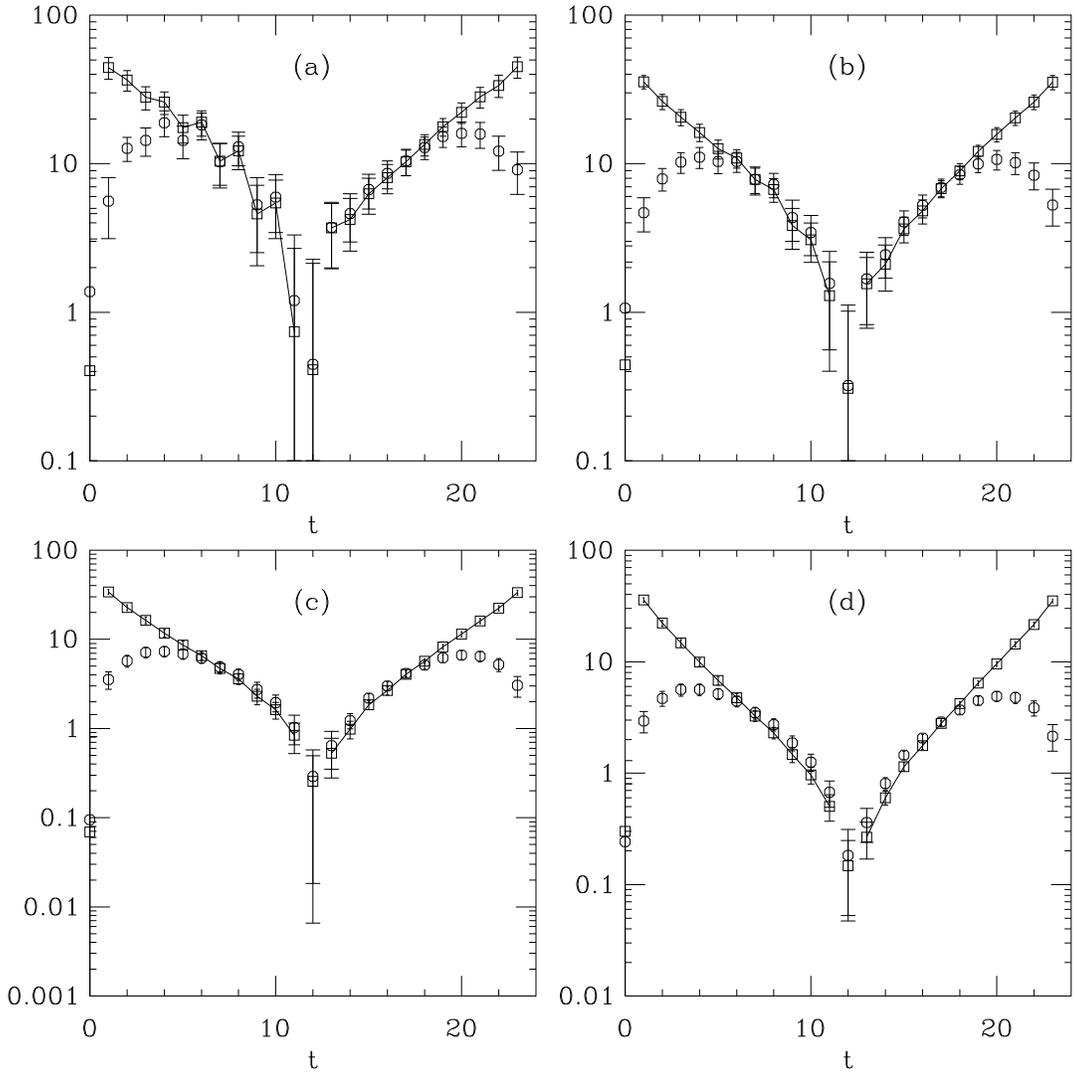}
\end{center}
\caption{
Comparison of the correlators needed to compute the PCAC quark mass
and $f_\pi$:
the (absolute value of the)
pseudoscalar source- axial vector sink correlators of the
  numerator of Eq. {\protect{\ref{RHO}}}.
 Squares (connected by lines) show the full hadron correlator.
Octagons show the contribution from the lowest 20 modes.
 (a) $am_q=0.01$ ($\pi/\rho\simeq 0.34$);
 (b) $am_q=0.02$ ($\pi/\rho\simeq 0.50$);
 (c) $am_q=0.04$ ($\pi/\rho\simeq 0.61$);
 (d) $am_q=0.06$ ($\pi/\rho\simeq 0.64$).}
\label{fig:compax}
\end{figure}

The numerators of the relevant correlators are shown in Fig. \ref{fig:compax}.
The source $C(0,0)$ is a Gaussian source ($\gamma_5)$. 
This picture plus Fig. \ref{fig:comppseudo}
 serves to show that  the average pseudoscalar correlator
is reproduced using only the lowest fermion modes in the quark propagator,
even at short $t$. By longer $t$, the axial current matrix element is also
 saturated by the low lying modes. Thus
both the PCAC quark mass and $f_\pi$ will be correctly computed using
 these truncated propagators at small
 quark mass--as a straightforward fit (see Table \ref{tab:mqfpi}) shows.
Note that by $am_q=0.06$ the 20-mode PCAC quark mass deviates from the
full calculation.

\begin{table}
\begin{tabular}{|c|l|l|l|l|}
\hline
$am_q$ & full $m_q$ & 20-mode $m_q$ & full $af_\pi$&  20 mode $f_\pi$ \\
\hline
   0.01 &  0.009(1) &  0.011(2)&  0.092(8) &  0.100(12)  \\
   0.02 &  0.020(1) &  0.017(2)&  0.091(4) &  0.100(6)  \\
   0.04 &  0.041(1) &  0.032(2)&  0.092(2) &  0.085(5)  \\
   0.06 &  0.062(1) &  0.046(2)&  0.093(2) &  0.085(10)\\
\hline
\end{tabular}
\caption{PCAC quark mass and pseudoscalar decay constant from
full propagators and from 20-mode truncations.}
\label{tab:mqfpi}
\end{table}

\subsection{Vector and Axial Vector Correlators}
\begin{figure}[thb]
\begin{center}
\epsfxsize=0.8 \hsize
\epsffile{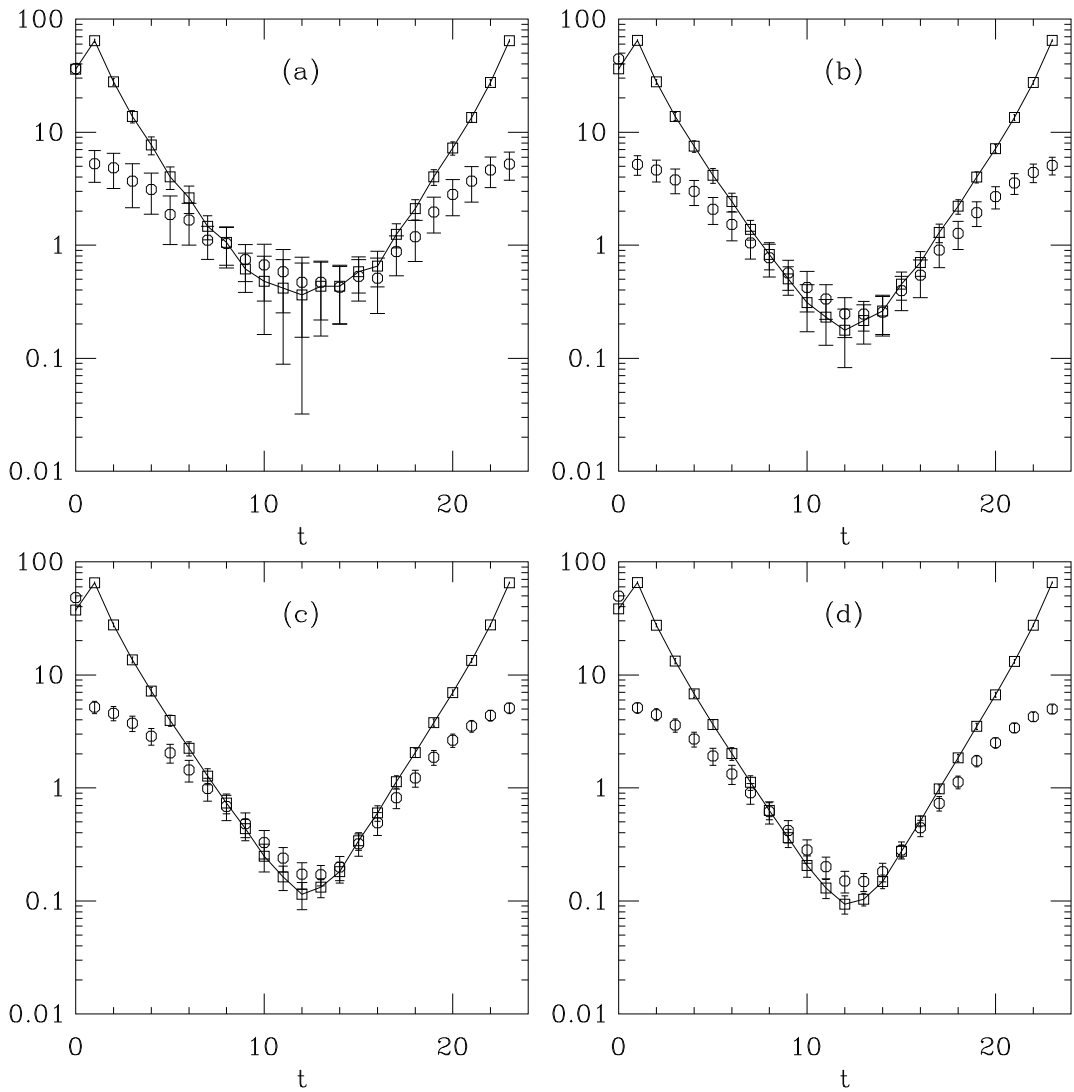}
\end{center}
\caption{
Comparison of the full   vector meson correlator (squares) with the correlator
built from the lowest  20 eigenmodes of $H(0)^2$ (octagons).
 (a) $am_q=0.01$ ($\pi/\rho\simeq 0.34$);
 (b) $am_q=0.02$ ($\pi/\rho\simeq 0.50$);
 (c) $am_q=0.04$ ($\pi/\rho\simeq 0.61$);
 (d) $am_q=0.06$ ($\pi/\rho\simeq 0.64$).
}
\label{fig:comprho}
\end{figure}

We observe that the vector meson correlator saturates only at a much larger
time separation than the pseudoscalar correlator. This is not very surprising
based on instanton model phenomenology\cite{ref:SS}. The two quarks of the
vector meson have to couple to two different instantons to propagate chirally.
That requires a propagation distance about twice the instanton size; at shorter
distances the quarks of the vector meson propagate like free particles,
independent of the instanton modes of the Dirac operator.
This is shown in Fig. \ref{fig:comprho}.

\begin{figure}[thb]
\begin{center}
\epsfxsize=0.4 \hsize
\epsffile{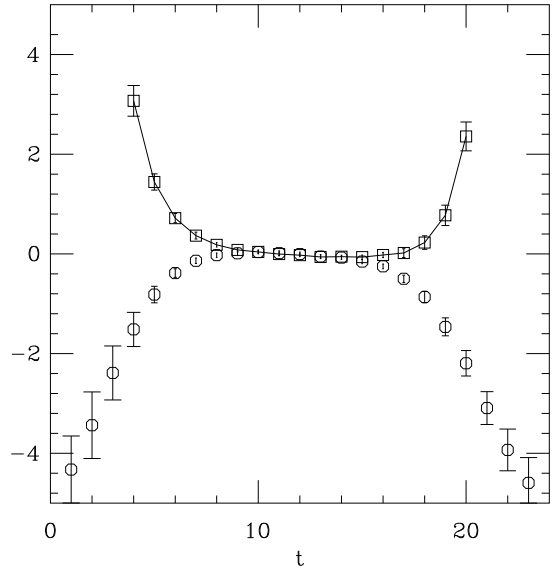}
\end{center}
\caption{
Comparison of the full  axial  vector meson correlator with the correlator
built from the lowest twenty eigenmodes of $H(0)^2$, at bare quark mass 0.04.
Squares are the full correlator; octagons, 20-eigenmode truncation.
}
\label{fig:comppv}
\end{figure}

The signal in the axial vector channel is much noisier, and if the mode sum
and the full propagators resemble each other, it is only after our
signal has disappeared into the noise.
At low $t$ the low mode correlator even has the opposite sign
to the full correlator. Compare Fig. \ref{fig:comppv}.

\subsection{Baryon Correlators}

\begin{figure}[thb]
\begin{center}
\epsfxsize=0.8 \hsize
\epsffile{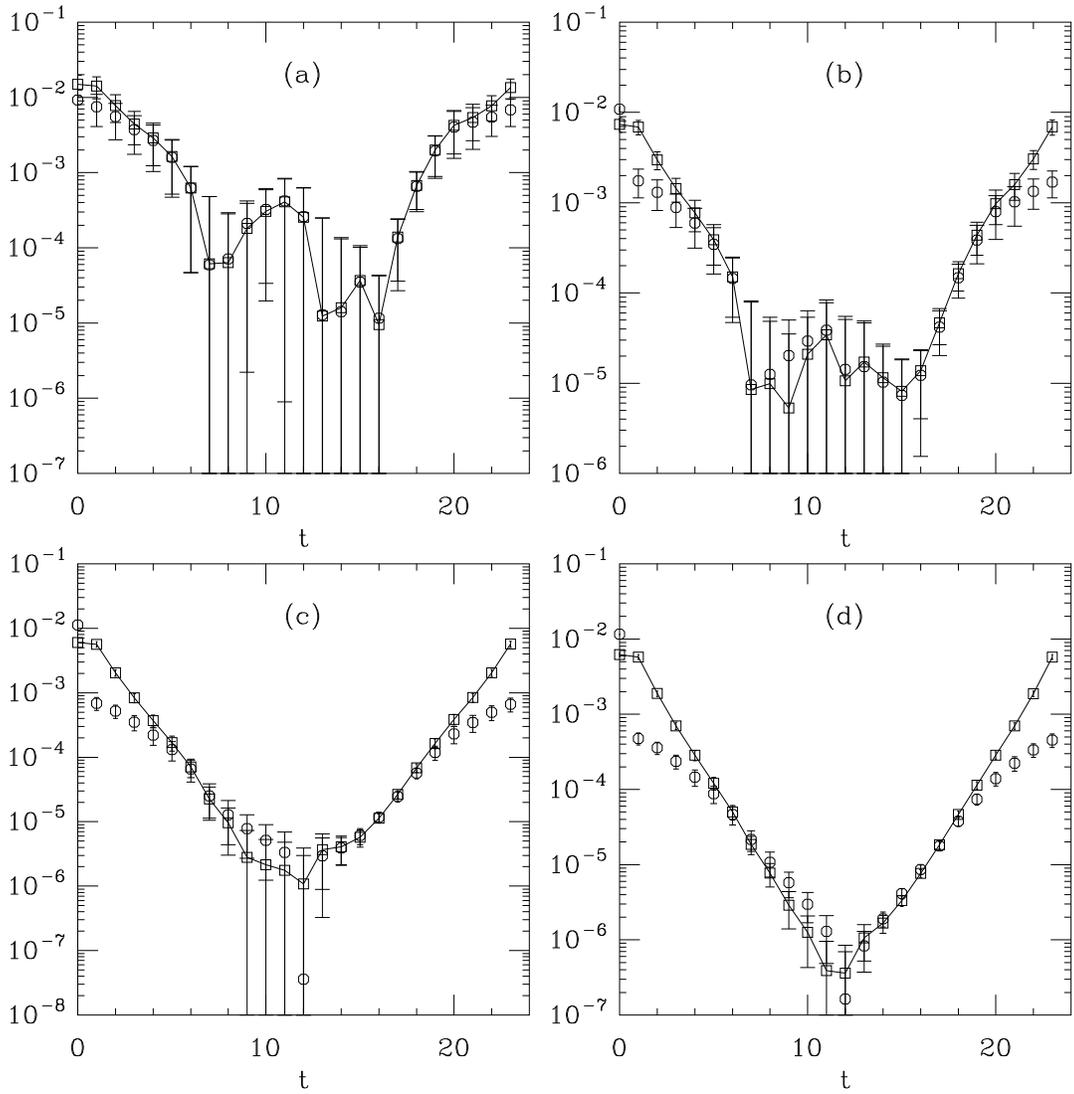}
\end{center}
\caption{
Comparison of the full nucleon  correlator (squares) with the correlator
built from the lowest  20 eigenmodes of $H(0)^2$ (octagons).
 (a) $am_q=0.01$ ($\pi/\rho\simeq 0.34$);
 (b) $am_q=0.02$ ($\pi/\rho\simeq 0.50$);
 (c) $am_q=0.04$ ($\pi/\rho\simeq 0.61$);
 (d) $am_q=0.06$ ($\pi/\rho\simeq 0.64$).
}
\label{fig:compprot}
\end{figure}

\begin{figure}[thb]
\begin{center}
\epsfxsize=0.8 \hsize
\epsffile{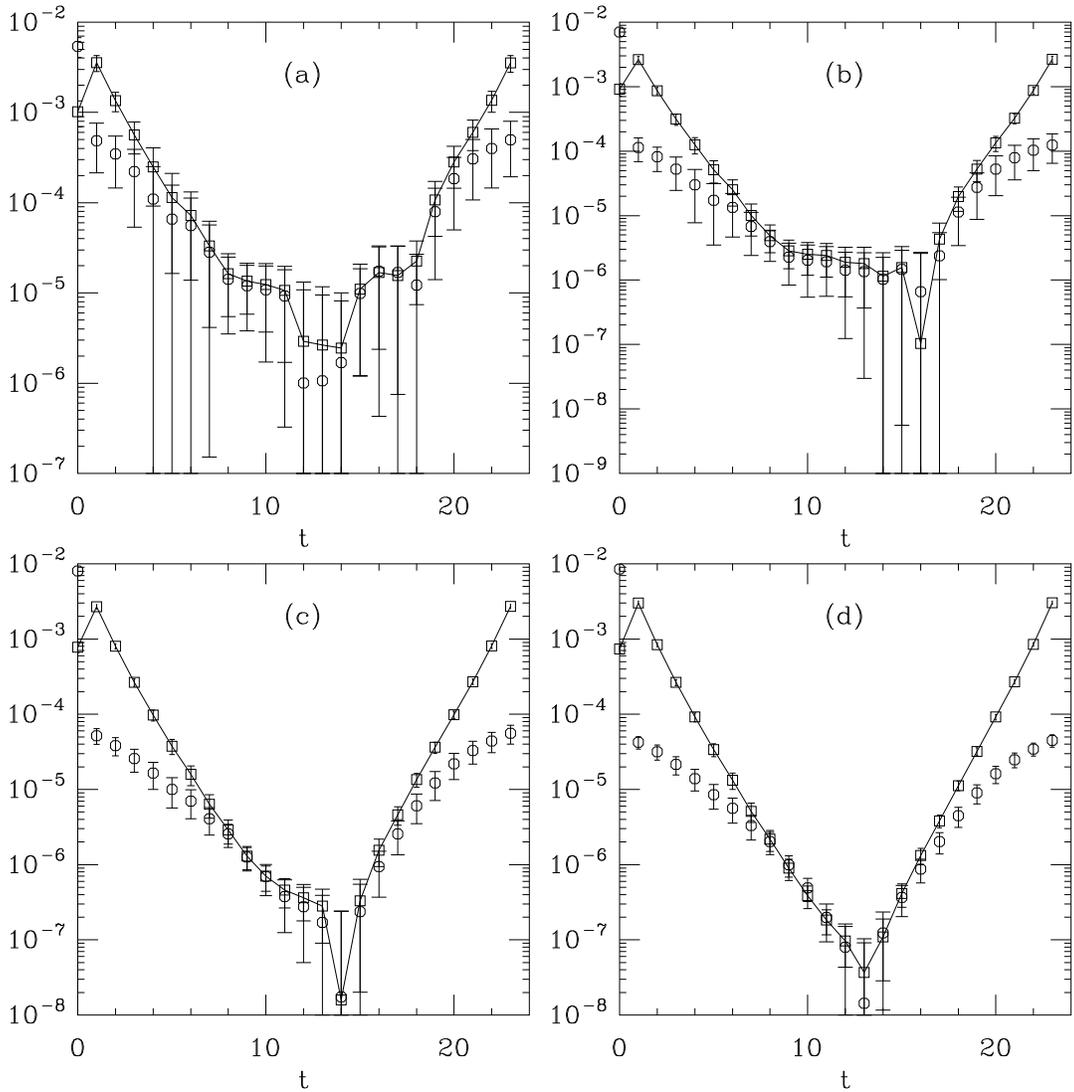}
\end{center}
\caption{
Comparison of the full delta correlator (squares) with the correlator
built from the lowest  20 eigenmodes of $H(0)^2$ (octagons).
 (a) $am_q=0.01$ ($\pi/\rho\simeq 0.34$);
 (b) $am_q=0.02$ ($\pi/\rho\simeq 0.50$);
 (c) $am_q=0.04$ ($\pi/\rho\simeq 0.61$);
 (d) $am_q=0.06$ ($\pi/\rho\simeq 0.64$).
}
\label{fig:compdelt}
\end{figure}

Both baryon signals (proton and delta) become increasingly noisy
 at small quark mass.  However, it appears that the low-lying fermionic
modes do a better job of saturating the nucleon correlator than the
delta correlator. Compare Figs. \ref{fig:compprot} and
\ref{fig:compdelt}.

These features are in complete agreement with simple instanton-based
phenomenology\cite{ref:SS}: In the instanton liquid, different flavor quarks
can travel together from instanton to instanton, exchanging their flavor
and flipping their spin. In pseudoscalar and scalar meson
channels, the quantum numbers of the quarks allow this kind of propagation.
Nucleons contain a spin-zero $ud$ diquark which  can also propagate in 
this way. 
The lower the value of the eigenmode, the more it couples to instantons,
and so the low eigenmodes dominate the correlator.
Vector mesons, however, lack first order instanton interactions
but do interact in second order.
The diquarks in the delta are all in $j=1$ states and cannot
 experience first-order interactions either.

According to the calculations in Ref. \cite{ref:SS}, the important
parameter of the instanton liquid phenomenology is the tunneling
amplitude between instantons and anti-instantons, 
\bee
\langle|T_{IA}|^2\rangle={{2\pi^2}\over{3N_c}}{N\over V}\rho^2
\ee
and with a mean instanton size of $\rho \simeq 1/3$ fm and density
$N/V \simeq 1$ fm${}^{-4}$, $|T_{IA}|\simeq 90$ MeV. The quark propagator
for quark mass $m$
in the zero-mode zone has a denominator of roughly $T+im$ and one might
 expect that
when $m$ becomes comparable to $T_{IA}$, replacing the full propagator
by a sum over instantons might be a bad approximation. A lattice regulated
bare quark mass of about 100 MeV would correspond roughly to $ma\simeq 0.06$,
which is actually the place where we observe that the lowest modes
begin not to
saturate any correlator. 

Our remarks only apply to long distance correlation functions,
not to short distance point-to-point correlators, nor to vacuum-to-vacuum
diagrams (hairpins), neither of which we have yet investigated.
Point-to-point correlators look at QCD at short to intermediate distances
and the results from these simulations cannot be compared directly with
ours. None the less, Ivanenko and Negele\cite{ref:IN}
 have tried to saturate the pseudoscalar 
and vector point to point correlators with low lying eigenmodes.
With 128 modes both correlators can be saturated.

Venkataraman and Kilcup\cite{ref:VK} studied
 long distance correlators on dynamical
lattices. They found that the pion correlator (at a pseudoscalar/vector meson
mass ratio of 0.55, from the data set of Ref. \cite{ref:Christ91}) could not
be saturated by the 32 lowest modes. They lattice volume is somewhat larger
than ours, and their action is less chiral which can explain why their findings
are different.

Several groups\cite{ref:VK,ref:Bardeen:2000cz} have reported that
pseudoscalar hairpins are saturated by low eigenmodes.

Finally, we\cite{ref:inst_only} and Kovacs\cite{ref:inst_onlyK}
 have tried to study the effects of instantons
in chiral symmetry breaking by isolating the instantons in a
 gauge configuration and reconstructing a pure
 multi-instanton configuration. Ordinary hadron spectroscopy using some
standard (nonchiral) action is then computed on these configurations.
 The vector channel (especially
as seen by Ref. \cite{ref:inst_onlyK}) is dominated by a very low
mass excitation even at higher quark  masses.
 These results qualitatively resemble the picture
of the low mode correlators
of Figs. \ref{fig:comppseudo} and \ref{fig:comprho}.

\section{Conclusions}
Instantons seem to be responsible for most of chiral symmetry breaking in
quenched QCD on the lattice, with a lattice spacing near 0.12 fm.
The lowest-eigenvalue eigenmodes of the Dirac operator have structure which
is strongly  correlated with the locations of instantons and
anti-instantons.
 Low-lying nonchiral modes  make a large contribution to
light quark mass hadron propagators
 in channels where instanton liquid phenomenology
would predict they would.
 Our little instanton liquid
calculation indicates that the instanton picture should break down
at larger quark masses, as we observe.

Of course, there is no reason for instantons to be the whole story.
While on
even smoother lattices one expects that instantons will be equally important,
in the strong coupling limit chiral symmetry is broken in QCD without
any recourse to instantons\cite{ref:OLD}. Some vestige of this
mechanism of chiral symmetry breaking might persist to the continuum limit.
Also, Fig. \ref{fig:drho} shows that the mean instanton size seen by our
fermions is about 0.3 fm. A larger lattice spacing will compromise these
objects. One might expect a different physical picture of chiral symmetry
breaking in lattice simulations in that case. 

Note also that instantons cannot account for the behavior of
the pion  at larger quark mass. From Fig. \ref{fig:mpi2mq}
we see that the linear relation between the squared pion mass and the
quark mass persists to well above a pseudoscalar/vector mass ratio of 0.8,
where the light modes we have identified no longer saturate the
 pion correlator.
 One would still say that the pion exhibits pseudo-Goldstone boson
behavior in spite of the fact that the low lying modes become
 progressively less important. One might imagine that more modes might
saturate the correlator.
However, the qualitative features of instanton liquid phenomenology
are supposed to involve only about as many modes as there are instantons,
so the need for more than a few modes in our simulation volume
begins to conflict with this phenomenology.

Nevertheless, in the real world, the up and down quarks are light.
Our work suggests strongly that instantons  affect their dynamics.

\section*{Acknowledgements}
Conversations with I.~Horvath and H.~Thacker
inspired this work. Conversations with J.~Negele and E.~Shuryak sharpened
our arguments.
We are deeply grateful for the design and construction of
our Beowulf cluster by  Doug Johnson.
 This work was supported by the
U.~S. Department of Energy.


\end{document}